\documentclass[conference]{IEEEtran}
\IEEEoverridecommandlockouts
\usepackage[utf8]{inputenc}

\usepackage{cite}
\usepackage{amsmath}
\usepackage{graphicx}
\usepackage{booktabs}
\usepackage{url}
\usepackage{xcolor}
\usepackage[hidelinks]{hyperref}
\def\BibTeX{{\rm B\kern-.05em{\sc i\kern-.025em b}\kern-.08em T\kern-.1667em\lower.7ex\hbox{E}\kern-.125emX}}

\newcommand{\pBplcc}{0.950}
\newcommand{\pBsrocc}{0.940}
\newcommand{\pBrmse}{0.36}
\newcommand{\pBn}{406}
\newcommand{\jBmzplcc}{0.870}
\newcommand{\jBmzsrocc}{0.874}

\newcommand{\jBbsplcc}{0.818}
\newcommand{\jBbssrocc}{0.793}

\newcommand{\vmMosplcc}{0.933}
\newcommand{\vmMossrocc}{0.931}
\newcommand{\vmMzplcc}{0.791}
\newcommand{\vmMzsrocc}{0.818}
\newcommand{\vmBsplcc}{0.822}
\newcommand{\vmBssrocc}{0.822}

\begin{document}

\title{JEVQA --- Video Quality from Metadata, Bitstream, and Pixel Features with a General-Purpose Decision Model}

\author{\IEEEauthorblockN{Werner Robitza}
\IEEEauthorblockA{AVEQ GmbH, Vienna, Austria\\
werner.robitza@aveq.info}}

\maketitle

\begin{abstract}
Instrumental quality models for video quality prediction are usually trained for a fixed set of codecs or other input features, and every new input variable requires retraining. Novel, general-purpose decision models can answer questions without task-specific training, but it is unclear whether they can judge video quality. We evaluate Jev, a commercial ``System One'' model that returns probability distributions over a provided answer scale, as a zero-shot video quality model. We call the resulting method JEVQA. In a first study on 1,936 AV1, H.264, HEVC, and VP9 encodes of 22 sources, scored against VMAF as ground truth, using encoding metadata only, JEVQA reached a Pearson correlation of 0.737, on par with the standardized ITU-T P.1204.1 model (0.733). Giving the model bitstream data raised the accuracy to 0.797, and combined pixel-based and bitstream features raised it to 0.824. A pixel-only variant failed in our tests. In a second study, using H.264, HEVC, and VP9 encodes in the AVT-VQDB-UHD-1 database, the metadata-only model reached a correlation of 0.879 with MOS, close to P.1204.1 (0.898). Bitstream statistics did not help there. Our results show that trained models on the same features remain clearly ahead in both studies, but that zero-shot classifiers are promising.
\end{abstract}


\section{Introduction}

Video streaming still accounts for the majority of consumer Internet traffic, and monitoring the quality customers receive is therefore important for network operators and content providers alike. Instrumental quality models are usually employed for this: they can estimate a Mean Opinion Score (MOS) or a comparable quality index from stream metadata, bitstream features, or decoded pixels. The ITU-T P.1203 and P.1204 series of Recommendations~\cite{itu-p1203,itu-p1204} standardize such models for several input levels, and the choice of input level is dictated by what the measurement system can see.

A fundamental limitation of these models is that they must be trained. Models are typically fitted to subjective datasets obtained with a fixed set of codecs and encoding settings. When a new codec is released (e.g., AV1), these models have to be retrained with new subjective data~\cite{Robitza2022,Rao2022}, or at least validated to be still accurate. At the same time, large pre-trained models have shown that many prediction tasks can be solved ``zero-shot'' without special training. Large multimodal models can score image and video quality from pixels when given the right instructions~\cite{Wu2024qbench,Wu2024qalign}, and text LLMs can be asked to rate encodes from their metadata~\cite{Goering2025llm}. However, these models are slow and often costly, making it hard to integrate them into 24/7 monitoring pipelines.

In this paper, we look at a different class of models. Jev~\cite{Almeida2026} is a commercial model that does not generate text tokens. It receives a structured state and a set of questions as input, and it returns, for each question, a calibrated probability distribution over a predefined set of ordered answers. Responses take fractions of a second and cost less than a cent, making the model more suitable for live tasks. However, whether such models are useful for video quality prediction is an open question.

We therefore wanted to find out how well a general-purpose decision model predicts video quality from the same kinds of inputs that standardized models receive, and how it compares with those models. This paper contributes: (1) JEVQA, a zero-shot video quality predictor built on Jev, and its evaluation against VMAF on 1,936 encodes across four codecs at four information levels---metadata only, pixel measurements, bitstream statistics, and combined pixel/bitstream data; (2) a validation of the metadata and bitstream levels against subjective scores on the 562 rated encodes of the AVT-VQDB-UHD-1 database~\cite{Rao2019a}; (3) a comparison with the ITU-T P.1204.1 Mode~0 and P.1204.3 full-bitstream models and with trained baselines on the same inputs, to quantify how much of the available signal the zero-shot model extracts; and (4) a discussion of the JEVQA state design.

The remainder of this paper is organized as follows: Section~\ref{sec:related} presents related work. Section~\ref{sec:method} describes the two datasets, the feature extraction, and how we queried the model. Results are presented in Section~\ref{sec:results}, with discussion and conclusion following in Sections~\ref{sec:discussion} and~\ref{sec:conclusion}.

\section{Related Work}
\label{sec:related}

Instrumental video quality models are commonly grouped by the information they may use. ITU-T P.1203~\cite{itu-p1203} defines four modes for HTTP adaptive streaming: Mode~0 uses only codec, bitrate, resolution, and frame rate; Mode~1 adds frame sizes and types; Mode~2 parses part of the bitstream; and Mode~3 parses the full bitstream, including quantization parameters (QP) and motion vectors. The P.1204 series~\cite{itu-p1204,Raake2020a} extends the scope to 4K, 60 fps, adds pixel-based and hybrid models, and support for HEVC and VP9. Ramachandra Rao et al.~\cite{Rao2022} generalized the bitstream approach into AVQBits, a family of models for each mode. Pixel-based no-reference models~\cite{Shahid2014,Goring2019b} instead compute features such as blockiness, blur, and noise from decoded frames and fit a regressor to MOS. A known difficulty with such features is their content dependence: spatial and temporal information of the source influences both perceived quality and clip compressibility~\cite{Robitza2021}. Deep-learning-based pixel models like Google's UVQ\footnote{\url{https://github.com/google/uvq}} show that predictions from pixel features alone---rather than engineered signal features---can be very accurate, at the cost of requiring more computation effort.

More recently, large multimodal models have been evaluated as quality judges: Q-Bench~\cite{Wu2024qbench} benchmarked foundation models on low-level vision questions, and Q-Align~\cite{Wu2024qalign} showed that answering with discrete text-defined levels (``excellent'' to ``bad'') produces useful image and video quality scores. Göring et al.~\cite{Goering2025llm} prompted 17 text-generating LLMs with a one-sentence description of codec, bitrate, frame rate, and resolution and asked for a 1--5 score. The best single model (DeepSeek-V3) reached a PLCC of 0.87. An ensemble of the best models achieved PLCC 0.90, close to the trained metadata model AVQBits$|$M0 (0.89) but far from P.1204.3 (0.97). The authors describe the approach as a proof of concept only, because the LLMs are slow.

Finally, VMAF~\cite{Li2016vmaf}, developed by Netflix, and recently updated to v1,\footnote{\url{https://netflixtechblog.com/vmaf-v1-good-is-not-good-enough-60d7e4244ea8}} is a full-reference metric fitted to subjective scores and widely used in the industry for quality-based encoding decisions and asset control. VMAF requires a video source, so it cannot be used in all monitoring environments. It is therefore the ground truth of our first study because it labels thousands of encodes repeatably at low cost; our second study validates the findings against MOS.

Almeida et al.~\cite{Almeida2026} released Jev in August 2026. As a ``System One'' type model (inspired by Daniel Kahneman's book \emph{Thinking, Fast and Slow}), it provides quick judgments for an input state, over predefined sets of output categories or ordinal scales. Instead of relying on human feedback for training, which they argue biases results, the authors used Reinforcement Learning for Calibrated Decisions (RLCD). Jev's claimed benefits are that the model is faster and cheaper to execute for large inputs when compared to typical LLMs (like ChatGPT or Claude).

In particular, considering Pinson's work on the accuracy and reliability of humans in subjective tests~\cite{pinson2023precision}, the development of models that can replace single human viewers and retain their detection accuracy is appealing for use cases where no trained models exist (yet).

Hence, given the current state of the art, we intend to discover whether Jev-type models can be useful for cheap and fast assessment of video quality.

\section{Methodology}
\label{sec:method}

\subsection{Study A: CRF Ladders Scored With VMAF}

We used the 22 source clips of the AOM Common Test Conditions class A2 set~\cite{aomctc}. They are 1080p clips (three in portrait orientation), 130 frames long, and a mix of 8-bit and 10-bit 4:2:0 content. From each source we created a ladder of 88 encodes with ffmpeg across four resolutions (1080p, 720p, 540p, and 360p short side) and four encoders with a constant rate factor (CRF) sweep: x264 (preset \emph{medium}) and x265 (preset \emph{fast}) at CRF 18--43 in steps of 5; libvpx-vp9 (\emph{good}, cpu-used 4) and SVT-AV1 (preset 8) at CRF 20--60 in steps of 10. This yields 1,936 processed video sequences (PVS). Each PVS was upscaled back to the source resolution with bicubic interpolation and scored with libvmaf using the VMAF v1.0.16 model for 1080p displays at three display heights viewing distance, with both inputs converted to 10-bit. The per-sequence ground truth is the arithmetic mean of the per-frame VMAF scores. For study A, encoding scripts and scores are available online.\footnote{\url{https://github.com/slhck/vmaf-v1-vs-v0}}

\subsection{Study B: AVT-VQDB-UHD-1 Scored With MOS}

AVT-VQDB-UHD-1~\cite{Rao2019a} is a public database of 4K sources encoded with x264, x265, and libvpx-vp9 in two-pass average-bitrate mode with 4:2:2 chroma subsampling, at 360p to 2160p and 60 frames per second, and rated by 24 to 29 participants per test on 65-inch and 55-inch 4K screens. We used its three codec tests (tests~1 to~3, 6 sources each, 8 to 10~s long). We explicitly excluded test~4, which varies the frame rate, since we wanted to avoid comparing fps vs. non-fps changes across different kinds of models. We also excluded the two test-2 encodes of the \emph{Dancers} source that the database paper reports as processing errors. This leaves 562 PVS with MOS on a 1--5 scale: 190 H.264, 216 HEVC, and 156 VP9 encodes.

The database includes VMAF scores, but they were computed with an older version and, for one source, against misaligned frames. We therefore fixed the alignment issues and recomputed VMAF v1.0.16 for every PVS with the 4K model at 1.5 display heights, matching the viewing setup of the test.

It must be noted that AVT-VQDB-UHD-1 was one of the databases used to train the P.1204-series models~\cite{Raake2020a}, so the standards-based models may have had an advantage here.

\subsection{Feature Extraction}

To obtain the input features for Jev, we used two analyzers: (1) The bitstream analyzer \emph{videoparser-ng}\footnote{\url{https://github.com/aveq-research/videoparser-ng}} parses H.264, HEVC, VP9, and AV1 streams and reports per-frame size, frame type, average QP, motion vector statistics, and coefficient statistics. In addition, it outputs the mode~0 data (codec, resolution, fps, bitrate) (2) Our proprietary tool \emph{video-analyzer} decodes the stream and reports per-frame no-reference measures: blockiness, blur, brightness, contrast, saturation, spatial and temporal information, high-frequency energy, noise, black and white levels, fine detail, and scene change indicators.

\subsection{JEVQA: The Decision Model and Its Inputs}

Jev answers typed questions about a JSON state. We used it via OpenRouter's Decisions API.\footnote{\url{https://openrouter.ai/typesafe/jev-1.13}} We therefore did not train or fine-tune the model; we only built the inputs for it. In the following we will call this approach JEVQA to keep it apart from the actual upstream model. To obtain the output, we used the \emph{score} question type, which returns a position and a probability distribution over an ordered set of answers. For correlating answers with VMAF, we chose ten VMAF bins as output (0--10, 10--20, \dots, 90--100), each labeled with an adjective from ``very poor'' to ``excellent''; for comparing with MOS output, we chose ten bins of the MOS range (1.0--1.4, \dots, 4.6--5.0) with the same adjectives, and, as a variant, the five labels of the Absolute Category Rating (ACR) scale. The prediction is the expected value over bin centers:
\begin{equation}
\hat{q} = \sum_{i=1}^{n} p_i \, c_i,
\end{equation}
where $p_i$ is the returned probability for bin $i$ and $c_i$ its center. We used the distribution mean rather than the most likely bin, which performed considerably worse in early runs.

We then defined four information levels. (1) \emph{Mode~0} mirrors the metadata-only input of standardized models, containing codec, bitrate in kbit/s, width, height, and frame rate. (2) The \emph{pixel level} contains only decoded-frame measurements plus the encoded width and height, without codec, bitrate, or QP, representing a measurement system without bitstream access. (3) The \emph{bitstream level} adds the \emph{videoparser-ng} output to the metadata, per frame. (4) The \emph{combined} level sends the bitstream level together with the pixel measurements. The model never received source names, the CRF or target bitrate setting, a reference video, or any explicit content descriptor.

For sequence-level prediction, the per-frame measurements were summarized as mean, standard deviation, minimum, and maximum, because sending every frame of a clip would exceed the API's token limit. Each state included a short guide describing only the features present in that request.

In our tests we iteratively improved the input state shapes over five versions (\emph{S1}--\emph{S5}): (1) S1 listed every frame's measurements and consequently returned API limit errors. (2) S2 used the raw per-codec quantizer values with a compact feature guide. (3) S3 added normalized quantizer values and motion fields, anchored the answer scale to VMAF reference points, and fixed per-frame columns that had wrongly been treated as sequence metadata. (4) S4 replaced all codec-specific quantizer values with a single H.264-equivalent 0--51 scale: H.264 and HEVC kept their QP (with the 10-bit H.264 offset of 12 removed), while the 0--255 quantizer index of VP9 and AV1 was multiplied by $51/255$. (5) S5 added a nonlinear quantizer-strength proxy, the 10th, 50th, and 90th percentiles of each feature, temporal-change statistics, and an optional ``decomposed'' question set with separate questions on overall quality, compression, detail, artifacts, and temporal quality. While S5 appears alongside S4 in the result tables, it did not improve accuracy, so our results will refer to S4 unless stated otherwise.

\subsection{Reference Models and Baselines}

We assessed JEVQA's accuracy against two standardized models on the same inputs. For Mode~0, we used ITU-T P.1204.1, fed with the same codec, bitrate, resolution, and frame rate values, in its standard PC condition with a 3840$\times$2160 display (and, in Study~A, a 1920$\times$1080 display as a sensitivity check). For the bitstream data, we used ITU-T P.1204.3 with the \emph{videoparser-ng}. As the standardized model does not support AV1, and the VP9 motion statistics were possibly unreliable in the \emph{videoparser-ng} implementation, the P.1204.3 comparisons used H.264 and HEVC only.

To quantify how much signal our feature tables actually contain, we additionally fitted baseline models on the same input features: A Ridge regression on bitrate alone, and on normalized QP plus bits per pixel, and tree ensembles (ExtraTrees, RandomForest, and HistGradientBoosting) on the full bitstream, pixel, and combined tables. For all those trained models we used leave-one-source-out cross-validation, keeping all encodes of a source clip in the same fold. We also fitted a Ridge model on the five distributions of the decomposed question set from S5. These supervised results will be reported separately from the zero-shot predictions.

\subsection{Evaluation}

Following ITU-T P.1401~\cite{itu-p1401}, we report the Pearson linear correlation coefficient (PLCC), the Spearman rank-order correlation coefficient (SROCC), and the root mean square error (RMSE). Because the reference models and JEVQA do not necessarily share the scale and offset of the ground truth, we report RMSE both raw and after a single database-level linear mapping (``RMSE lin.'').

For frame-level runs, we evaluate the pooled correlation across all frames and, separately, the median correlation within each PVS, since a model may separate clips well while not tracking quality within a clip.

We further conducted pairwise tests: we asked the model to compare two adjacent encodes of the same ladder (same source, resolution, and codec; neighboring CRF for the AOM CTC dataset, neighboring target bitrate for AVT-VQDB-UHD-1), and reported pair accuracy (proportion of correcly identified pairs), the Brier score (mean squared error of the returned probability against the correct answer; lower is better), and the median within-ladder SROCC. The pairwise test used the S5 state in Study~A and the S4 state in Study~B. For the input state, we alternated the candidate order to prevent a position shortcut.

\section{Results}
\label{sec:results}

\subsection{Study A: Sequence-Level Prediction by Information Level}

\begin{figure*}[t]
\centering
\includegraphics[width=\textwidth]{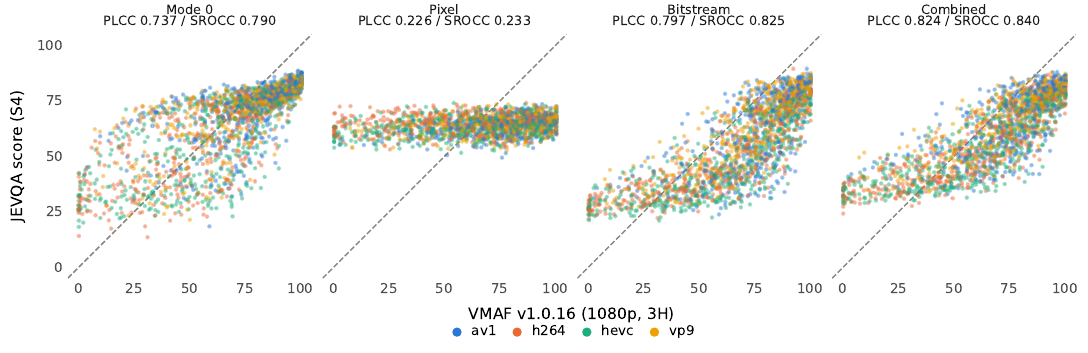}
\caption{Study A: JEVQA score against VMAF for the four information levels, one point per PVS, colored by codec.}
\label{fig:scatter}
\end{figure*}

Table~\ref{tab:main} gives the sequence-level results for all 1,936 PVS, and Fig.~\ref{fig:scatter} shows the corresponding scatter plots. With metadata only, JEVQA reaches a PLCC of 0.737 and an SROCC of 0.790, despite the state containing only five numbers and no content information. Adding bitstream statistics raises PLCC by 0.060 and SROCC by 0.035; further adding pixel measurements yields the highest zero-shot correlation (PLCC 0.824, SROCC 0.840). The pixel-only state, however, is close to uninformative (PLCC 0.226). Fig.~\ref{fig:scatter} also shows that all successful states compress the prediction range: predictions rarely exceed 88 even where VMAF reaches 100, which explains the gap between the raw and the mapped RMSE. The S5 state with decomposed questions did not improve on S4 (PLCC 0.815 vs. 0.824 for the combined level). For comparison, a supervised Ridge fit over the decomposed distributions reaches PLCC 0.850.

\begin{table}[t]
\caption{Study A: sequence-level results of the JEVQA states on all 1,936 PVS against VMAF.}
\label{tab:main}
\centering
\setlength{\tabcolsep}{3.5pt}
\footnotesize
\begin{tabular}{llrrrr}
\toprule
State & Level & PLCC & SROCC & RMSE & RMSE lin. \\
\midrule
S4 & Mode 0 & 0.737 & 0.790 & 17.55 & 17.27 \\
S4 & Pixel & 0.226 & 0.233 & 25.15 & 24.88 \\
S4 & Bitstream & 0.797 & 0.825 & 19.50 & 15.43 \\
S4 & Combined & \textbf{0.824} & \textbf{0.840} & \textbf{16.63} & \textbf{14.46} \\
S5 decomposed & Combined & 0.815 & 0.828 & 17.83 & 14.79 \\
\midrule
S5 dec.\ + Ridge & Combined & 0.850 & 0.856 & 13.47 & 13.45 \\
\bottomrule
\end{tabular}
\end{table}

The state versions had a larger effect than the information level. For the bitstream level, PLCC rose from 0.576 (S2) over 0.641 (S3) to 0.797 (S4). The S2 state, which sent raw quantizer values, reached PLCC 0.82 for H.264 and HEVC but only 0.30 for AV1 and 0.19 for VP9: the model read a VP9 quantizer index of 150 as if it were on the H.264 scale, which ends at 51, and predicted low quality throughout. Mapping all quantizers to one 0--51 scale in S4 brought AV1 to 0.69 and VP9 to 0.83 without changing the H.264 and HEVC results. We did not find another single change with an effect of this size.

\subsection{Study A: Comparison With P.1204 and Trained Baselines}

Table~\ref{tab:p12041} compares JEVQA Mode~0 with the P.1204.1 Mode~0 model on all PVS. JEVQA is slightly ahead on all three metrics, but the difference is small (PLCC 0.737 vs. 0.733). Per codec, P.1204.1 has the higher PLCC for AV1 (0.626 vs. 0.587 for JEVQA) and for VP9 (0.755 vs. 0.707). JEVQA is ahead for H.264 (0.773 vs. 0.765) and for HEVC (0.724 vs. 0.711). The data therefore do not support a claim that either Mode~0 model dominates the other. For both, AV1 is the hardest codec, which we attribute to specific encoding choices made with SVT-AV1 at preset~8 on this content. The two Ridge fits in Table~\ref{tab:p12041} are trained baselines. Bitrate alone, fitted to VMAF, explains less variance (PLCC 0.433) than JEVQA extracts from bitrate, codec, and resolution. A two-feature model with normalized QP and bits per pixel, which needs bitstream access, reaches PLCC 0.742. The zero-shot metadata model thus matches a small trained model that additionally sees the quantizer.

\begin{table}[t]
\caption{Study A: JEVQA Mode 0 against ITU-T P.1204.1 Mode 0 and two trained Ridge baselines on all 1,936 PVS.}
\label{tab:p12041}
\centering
\small
\begin{tabular}{lrrr}
\toprule
Model & PLCC & SROCC & RMSE lin. \\
\midrule
JEVQA S4 Mode 0 & \textbf{0.737} & \textbf{0.790} & \textbf{17.27} \\
P.1204.1, PC 3840$\times$2160 & 0.733 & 0.772 & 17.37 \\
P.1204.1, PC 1920$\times$1080 & 0.721 & 0.781 & 17.70 \\
\midrule
Ridge, bitrate only & 0.433 & 0.676 & 23.03 \\
Ridge, norm. QP + bits/pixel & 0.742 & 0.769 & 17.12 \\
\bottomrule
\end{tabular}
\end{table}

Table~\ref{tab:p12043} compares the models on the 1,056 H.264 and HEVC PVS that the P.1204.3 implementation supports. P.1204.3 performs best and is well ahead of JEVQA with bitstream input and with the combined state. JEVQA Mode~0 and P.1204.1 repeat the pattern of the full dataset. For comparison, on bitstream features, the ExtraTrees model fitted directly to VMAF with leave-one-source-out validation reaches PLCC 0.957 and RMSE 7.5 on all 1,936 PVS. The HistGradientBoosting model reaches 0.932 at frame level, showing that the bitstream tables contain substantially more information than JEVQA extracts. The pixel features are also informative when trained (PLCC 0.596 with the RandomForest model), but their ceiling is much lower on this dataset, and adding them to the trained bitstream model did not improve it.

\begin{table}[t]
\caption{Study A: results on the 1,056 H.264 and HEVC PVS supported by the P.1204.3 implementation.}
\label{tab:p12043}
\centering
\setlength{\tabcolsep}{3pt}
\footnotesize
\begin{tabular}{llrrr}
\toprule
Model & Level & PLCC & SROCC & RMSE lin. \\
\midrule
JEVQA S4 & Mode 0 & 0.747 & 0.792 & 18.38 \\
P.1204.1, PC 3840$\times$2160 & Mode 0 & 0.737 & 0.770 & 18.68 \\
JEVQA S4 & Bitstream & 0.814 & 0.852 & 16.06 \\
JEVQA S4 & Combined & 0.839 & 0.865 & 15.04 \\
P.1204.3 & Bitstream & \textbf{0.888} & \textbf{0.936} & \textbf{12.71} \\
\bottomrule
\end{tabular}
\end{table}

\subsection{Study A: Frame-Level and Pairwise Prediction}

Since we had per-frame features, we wanted to check whether JEVQA can track quality over time. When evaluated on every eighth frame of 352 PVS from four sources with bitstream features (5,792 requests), the S4 state reaches a pooled PLCC of 0.800, comparable to the sequence level. Within an individual PVS, however, the median PLCC is only 0.32 and the median SROCC 0.16. Therefore, pooling allows us to separate PVSs from one another, but the model does not follow frame-to-frame variations like VMAF can.

Table~\ref{tab:pairwise} reports the pairwise test over 288 adjacent CRF pairs from 64 ladders. With bitstream features, the model ranks every pair correctly with near-certain probabilities (Brier score 0.0001), and the combined state is almost as good. We can see that the pixel state is mostly wrong---inversed, in fact: it picks the lower-quality encode in 78\% of pairs.

\begin{table}[t]
\caption{Pairwise comparison of adjacent encodes within source/resolution/codec ladders.}
\label{tab:pairwise}
\centering
\small
\setlength{\tabcolsep}{4pt}
\begin{tabular}{llrrr}
\toprule
Study & Level & Pair acc. & Brier & Median SROCC \\
\midrule
A & Bitstream & \textbf{1.000} & \textbf{0.0001} & \textbf{1.000} \\
A & Pixel & 0.222 & 0.582 & $-0.893$ \\
A & Combined & 0.997 & 0.012 & 1.000 \\
B & Bitstream & 0.879 & 0.116 & 1.000 \\
\bottomrule
\end{tabular}
\end{table}

\subsection{Study B: Subjective Validation on AVT-VQDB-UHD-1}

\begin{figure}[t]
\centering
\includegraphics[width=\columnwidth]{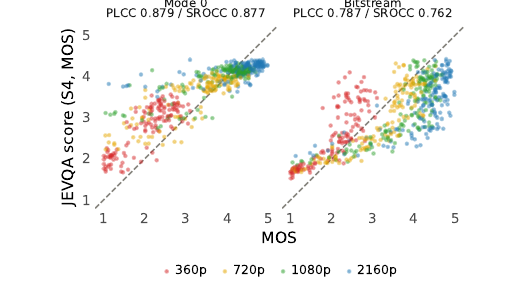}
\caption{Study B: JEVQA score against MOS on AVT-VQDB-UHD-1, colored by resolution.}
\label{fig:avt}
\vspace{-.5cm}
\end{figure}

Table~\ref{tab:avt} gives the results against MOS for all 562 PVS, and Fig.~\ref{fig:avt} shows the scatter plots. With metadata only, JEVQA reaches a remarkable PLCC of 0.879 and an SROCC of 0.877 against MOS, higher than against VMAF in Study~A, and close to P.1204.1 (0.898 and 0.890), which was trained on this database.

Asking Jev for an ACR label instead of ten MOS bins only resulted in small improvements (PLCC of 0.884, SROCC of 0.880). Per codec, the metadata model is weakest for H.264 (PLCC 0.831) and reaches 0.914 for HEVC and 0.904 for VP9; P.1204.1 shows the same order (0.848, 0.916, 0.939). Per resolution, both models separate encodes well at 360p to 1080p (PLCC 0.78 to 0.83 for JEVQA) but less well at 2160p (0.62), where the MOS range is narrow. On test~1 alone, the 180 PVS used by Göring et al.~\cite{Goering2025llm} (ten of the 190 H.264 encodes belong to other tests), JEVQA Mode~0 reaches PLCC 0.873 (ten bins) and 0.881 (ACR labels) with SROCC 0.857 and 0.862. This matches Göring's best single LLM (PLCC 0.870, SROCC 0.877) and stays slightly below their trained reference model (PLCC 0.891) and their best LLM ensemble (0.902).

\begin{table}[t]
\caption{Study B: results against MOS on AVT-VQDB-UHD-1.}
\label{tab:avt}
\centering
\setlength{\tabcolsep}{2.5pt}
\footnotesize
\begin{tabular}{llrrr}
\toprule
Model & Level & PLCC & SROCC & RMSE lin. \\
\midrule
JEVQA S4, 10 bins & Mode 0 & 0.879 & 0.877 & 0.53 \\
JEVQA S4, ACR & Mode 0 & 0.884 & 0.880 & 0.52 \\
JEVQA S4, 10 bins & Bitstream & 0.787 & 0.763 & 0.69 \\
JEVQA S4, ACR & Bitstream & 0.750 & 0.736 & 0.74 \\
JEVQA S5 decomposed & Bitstream & 0.754 & 0.742 & 0.74 \\
\midrule
P.1204.1 (in-sample) & Mode 0 & \textbf{0.898} & \textbf{0.890} & \textbf{0.49} \\
P.1204.3, \pBn{} H.264/HEVC & Bitstream & \pBplcc{} & \pBsrocc{} & \pBrmse{} \\
VMAF v1 4K (full-ref.) & Pixel & \vmMosplcc{} & \vmMossrocc{} & -- \\
\midrule
S5 dec.\ + Ridge & Bitstream & 0.872 & 0.848 & 0.55 \\
Ridge, bitrate only & Bitstream & 0.534 & 0.749 & 0.95 \\
ExtraTrees & Bitstream & 0.951 & 0.925 & 0.35 \\
\bottomrule
\end{tabular}
\end{table}

The bitstream level, which improved accuracy in Study~A, degraded performance in Study~B: PLCC drops to 0.787 with the ten-bin scale and to 0.750 with ACR labels. S5 did not help (0.754). As Fig.~\ref{fig:avt} shows, within each resolution, the bitstream state performs as well as the metadata state (PLCC 0.82, 0.88, 0.81, and 0.70 for 360p, 720p, 1080p, and 2160p, compared to 0.83, 0.83, 0.78, and 0.62 for Mode~0), but it misplaces resolutions relative to one another: its mean prediction is 0.88 MOS too low for 2160p encodes and 0.53 too high for 360p encodes, whereas the metadata state overrates 360p by 0.73 and is nearly unbiased at 2160p. The pairwise test in Table~\ref{tab:pairwise} confirms that ordinal judgment within a ladder is intact: over 371 adjacent bitrate pairs from 174 ladders, the model picks the better-rated encode in 88\% of cases with a median within-ladder SROCC of 1.0. The trained ExtraTrees model on the same bitstream features reaches PLCC 0.951, and the Ridge fit over the decomposed S5 outputs 0.872, showing that the feature tables \emph{do} contain the necessary signal; the zero-shot state simply fails to calibrate scores across resolutions.

On the \pBn{} H.264 and HEVC PVS supported by the P.1204.3 implementation, that model reaches PLCC \pBplcc{} and SROCC \pBsrocc{} against MOS, compared with \jBmzplcc{} and \jBmzsrocc{} for JEVQA Mode~0 and \jBbsplcc{} and \jBbssrocc{} for JEVQA with bitstream input on the same subset. The recomputed VMAF, a full-reference metric, correlates with MOS at PLCC \vmMosplcc{} and SROCC \vmMossrocc{} on all 562 PVS. Scoring the same JEVQA predictions against this VMAF instead of MOS gives PLCC \vmMzplcc{} and SROCC \vmMzsrocc{} for Mode~0 and \vmBsplcc{} and \vmBssrocc{} for the bitstream level. The metadata-only prediction thus agrees more with the subjective ratings than with the full-reference metric, and the gap between the two input levels disappears against VMAF.

\section{Discussion}
\label{sec:discussion}

\subsection{What Does the Model Know About Video Quality?}

JEVQA matches P.1204.1 against VMAF on simple encoding ladders, and is almost as accurate on a database that P.1204.1 was partly trained on. It is on par with text-based LLMs~\cite{Goering2025llm} at a fraction of their latency and cost. We can conclude that the model knows that 1~Mbit/s is too little for 1080p H.264 but enough for 360p, and that VP9 and AV1 reach the same quality at lower bitrates.

Bitstream features as input yielded mixed results. In Study~A, the model was given QP but did not outperform trained models: P.1204.3 was 0.07 PLCC ahead on the same subset, and a tree ensemble was 0.16 ahead. In Study~B, bitstream statistics made the prediction even worse. We assume that Jev interpreted a low QP as high quality, regardless of resolution. This is true for CRF-based encodes, but not for bitrate-controlled encodes, where QP follows rate control. In other words, on a 4K screen, a 360p encode at 1~Mbit/s has a low QP and a low MOS, a 2160p encode at 40~Mbit/s has a higher QP and a high MOS. Trained models would typically learn this interaction. The P.1204-based models even explicitly model resolution and quantization degradation separately and therefore do not have this failure mode. A zero-shot model must infer this from feature names and a short guide. The pairwise results fit this observation: within a resolution ladder, the model correctly rates the direction, but it lacks the calibration across conditions needed for absolute scores.

\subsection{Why Did Pixel States Fail?}

In Study~A, the pixel-only state actually inverted the quality ranking. The reason seems to be that as resolutions increase, high-frequency energy, spatial and temporal information, and noise also increase, because heavier compression removes them. A model can get the direction wrong if it understands noise or high-frequency energy alone as a sign of degradation. The trained pixel baseline (PLCC 0.596), however, shows that the features carry relevant signal, but a zero-shot classifier does not know the meaning of each feature in this context. Again, this is a state-design problem rather than evidence that pixel features are useless. Our relatively low baseline accuracy of 0.596 further shows that the chosen no-reference pixel measures are likely not reaching bitstream accuracy. We believe that this is why more modern approaches to no-reference pixel-based model development using deep learning can achieve more favorable results, in general.

\subsection{Limitations}

First, Study~A used VMAF rather than MOS as ground truth, and we know VMAF does not perfectly replace MOS. Second, the P.1204 models were partly trained on AVT-VQDB-UHD-1, so Study~B compared zero-shot predictions against in-sample references; a fair comparison would need a database unseen by either side. Third, neither dataset covered longer sessions containing adaptive streaming artifacts like initial loading delay, stalling, and quality variation. Fourth, no user-generated content was tested for the pixel-based features. Fifth, Jev is a commercial service with closed weights; later releases may behave differently from the version we recorded, and alternatives are likely to be released soon. Sixth, the P.1204.3 comparison covered only H.264 and HEVC, a limit of the reference implementation that we plan to address.

\section{Conclusion and Future Work}
\label{sec:conclusion}

We evaluated Jev, a general-purpose decision model, as a zero-shot video quality predictor on 1,936 encodes across four codecs scored with VMAF, and on 562 subjectively rated encodes of AVT-VQDB-UHD-1. JEVQA matched the ITU-T P.1204.1 Mode~0 model against VMAF and came close to it against MOS. With bitstream statistics, it ranked adjacent encodes correctly in both studies and improved VMAF prediction on CRF-controlled encodes, but on bitrate-controlled encodes it misplaced resolutions and fell behind P.1204.1. Trained models were still the most accurate on the same features. Jev failed on pixel features alone, and frame-level tracking did not work. We conclude that zero-shot predictors can be useful, cheap and fast, but that feature and input state design matter most for successful practical use. All Jev requests of both studies together, including every state version, frame-level, and pairwise run, cost about \$4; the Mode~0 run over all 1,936 PVS of Study~A completed in 39~s.

Jev-like models may be most useful where no instrumental model exists yet, for new codecs or as a first estimate before a subjective test. Future work will include improvements for the raw pixel and bitstream features, enhancing the input state representations, and adding more tests on subjective databases that the standardized models have not seen. Lastly, we want to also conduct more statistical analyses with respect to Jev's ability to replace one human viewer.

\section*{Acknowledgements}

During the preparation of this work, the author used Claude Fable 5.1 and OpenAI GPT-5.6 Sol in order to program analysis and plotting scripts, and cross-check the output data. The author reviewed and edited the content as needed and takes full responsibility for the content of the publication. The manuscript was copyedited using Gemini 3.7 Flash.

\bibliographystyle{IEEEtran}
\bibliography{refs}

\end{document}